\documentstyle [12pt] {article}
\begin{document}
\def\be{\begin{eqnarray}}
\def\en{\end{eqnarray}}
\def\non{\nonumber}
\def\la{\langle}
\def\ra{\rangle}
\def\ep{\varepsilon}
\def\b{\Lambda_b\to J/\psi\Lambda}
\def\j{{J/\psi}}
\def\Ri{\Rightarrow}
\def\bb{{\bar{B}}}
\def\pr{{\sl Phys. Rev.}~}
\def\prl{{\sl Phys. Rev. Lett.}~}
\def\pl{{\sl Phys. Lett.}~}
\def\np{{\sl Nucl. Phys.}~}
\def\zp{{\sl Z. Phys.}~}

\font\el=cmbx10 scaled \magstep2
{\obeylines
\hfill IP-ASTP-04-95
\hfill March, 1995}

\vskip 1.5 cm

\centerline{\large\bf Hadronic Weak Decays of Heavy Mesons and
Nonfactorization}
\medskip
\bigskip
\medskip
\centerline{\bf Hai-Yang Cheng}
\medskip
\centerline{Institute of Physics, Academia Sinica}
\centerline{Taipei, Taiwan 11529, Republic of China}
\bigskip
\bigskip
\bigskip
\centerline{\bf Abstract}
\bigskip
{\small
The parameters $\chi_{1,2}$, which measure nonfactorizable soft gluon
contributions to hadronic weak decays of mesons, are updated by extracting
them from the data of $D,\,B\to PP,~VP$ decays ($P$: pseudoscalar meson, $V$:
vector meson). It is found that $\chi_2$ ranges from $-0.36$ to $-0.60$ in the
decays from $D\to\bar{K}\pi$ to $D^+\to\phi\pi^+,~D\to\bar{K}^*\pi$, while
it is of order 10\% with a positive sign in $B\to\psi K,~D\pi,~D^*\pi,~D\rho$
decays. Therefore, the effective parameter $a_2$ is process dependent in charm
decay, whereas it stays fairly stable in $B$ decay. This implies the picture
that nonfactorizable effects become stronger when the decay particles become
less energetic after hadronization. As for $D,\,B\to VV$ decays, the presence
of nonfactorizable terms in general prevents a possible definition of
effective $a_1$ and $a_2$. This is reenforced by the observation of a large
longitudinal polarization fraction in $B\to\psi K^*$ decay, implying $S$-wave
dominated nonfactorizable effects. The nonfactorizable term dominated by the
$S$-wave is also essential for understanding the decay rate of $B^-\to D^{*0}
\rho^-$. It is found that all nonfactorizable effects  $A_1^{nf}/A_1^{BK^*},~A
_1^{nf}/A_1^{B\rho},~A_1^{nf}/A_1^{BD^*}$ ($nf$ standing for nonfactorization)
are positive and of order 10\%, in accordance with $\chi_2(B\to D(D^*)\pi(
\rho))$ and $\chi_2(B\to\psi K)$. However, we show that in $D\to\bar{K}^*\rho$
decay nonfactorizable effects cannot be dominated by the $S$-wave. A
polarization measurement in the color- and Cabibbo-suppressed decay mode
$D^+\to\phi\rho^+$ is strongly urged in order to test if $A_2^{nf}/A_2$ plays
a more pivotal role than $A_1^{nf}/A_1$ in charm decay.
}

\pagebreak
\noindent {\bf 1. Introduction}

It is customary to assume that two-body nonleptonic weak decays of heavy mesons
are dominated by factorizable contributions. Under this assumption, the
spectator meson decay amplitude is the product of the universal parameter
$a_1$ (for external $W$-emission) or $a_2$ (for internal $W$-emission),
which is channel independent in $D$ or $B$ decays, and hadronic matrix
elements which can be factorized as the product of two independent hadronic
currents. The universal parameters $a_1$ and $a_2$ are
related to the Wilson coefficient functions $c_1$ and $c_2$ by
\be
a_1=\,c_1+{1\over N_c}c_2,~~~~~a_2=\,c_2+{1\over N_c}c_1,
\en
with $N_c$ being the number of colors. It is known that the bulk of
exclusive nonleptonic charm decay data cannot be explained by this
factorization approach [1]. For example, the predicted ratio of the
color-suppressed mode $D^0\to \bar{K}^0\pi^0$ and color-favored decay
$D^0\to K^-\pi^+$ is in violent disagreement with
experiment. This signals the importance of the nonfactorizable effects.

  The leading nonfactorizable contribution arises from the soft gluon exchange
between two color-octet currents
\be
O_c=\,{1\over 2}(\bar{q}_1\lambda^a q_2)(\bar{q}_3\lambda^aq_4),
\en
where $(\bar{q}_1\lambda^a q_2)$ stands for $\bar{q}_1\gamma_\mu(1-\gamma_5)
\lambda^a q_2$. For $M\to PP,~VP$ decays ($P$: pseudoscalar meson, $V$: vector
meson), the nonfactorizable effect amounts to a redefinition of the
parameters $a_1$ and $a_2$ [2],
\footnote{Note that our definition of $\chi_1$ and $\chi_2$ is different
from $r_1$ and $r_2$ defined in [3] by a factor of 2.}
\be
a_1\to\,c_1+c_2({1\over N_c}+\chi_1),~~~~a_2\to\,c_2+c_1({1\over N_c}+\chi_2),
\en
where $\chi_1$ and $\chi_2$ denote the contributions of $O_c$ to color-favored
and color-suppressed decay amplitudes respectively relative to the
factorizable ones. For example, for $D_s^+\to\phi\pi^+,~D^+\to\phi\pi^+$
decays,
\be
\chi_1(D_s^+\to\phi\pi^+) &=& {\la\phi\pi^+|{1\over 2}(\bar{u}\lambda^a d)
(\bar{s}\lambda^ac)|D_s^+\ra  \over\la\phi\pi^+|(\bar{u}d)(\bar{s}c)|D_s^+
\ra_f },\non \\
\chi_2(D^+\to\phi\pi^+) &=& {\la\phi\pi^+|{1\over 2}(\bar{u}\lambda^a c)(\bar
{s}\lambda^as)|D^+\ra  \over\la\phi\pi^+|(\bar{u}c)(\bar{s}s)|D^+\ra_f }.
\en
The subscript $f$ in Eq.(4) denotes a factorizable contribution:
\be
\la\phi\pi^+|(\bar{u}d)(\bar{s}c)|D_s^+\ra_f &=& 2m_\phi f_\pi(\ep^*\cdot p_{_
{D_s}})A_0^{D_s\phi}(m_\pi^2),   \non \\
\la\phi\pi^+|(\bar{u}c)(\bar{s}s)|D^+\ra_f &=& m_\phi f_\phi(\ep^*\cdot p_{_
{D}})F_1^{D\pi}(m_\phi^2),
\en
where $\ep_\mu$ is the polarization vector of the $\phi$ meson, and
we have followed Ref.[4] for the definition of form factors. The
nonfactorizable contributions have the expressions
\be
\la\phi\pi^+|{1\over 2}(\bar{u}\lambda^ad)(\bar{s}\lambda^ac)|D_s^+\ra &=&
2m_\phi f_\pi(\ep^*\cdot p_{_{D_s}})A_0^{nf}(m_\pi^2),   \non \\
\la\phi\pi^+|{1\over 2}(\bar{u}\lambda^ac)(\bar{s}\lambda^as)|D^+\ra &=&
m_\phi f_\phi(\ep^*\cdot p_{_{D}})F_1^{nf}(m_\phi^2),
\en
with the superscript $nf$ referring to nonfactorizable contributions. It is
clear that
\be
\chi_1(D_s^+\to\phi\pi^+)=\,{A_0^{nf}(m_\pi^2)\over A_0^{D_s\phi}(m_\pi^2)},
{}~~~~\chi_2(D^+\to\phi\pi^+)=\,{F_1^{nf}(m_\phi^2)\over F_1^{D\pi}(m_\phi^2)}.
\en
That is, $\chi$ simply measures the fraction of nonfactorizable
contributions to the form factor under consideration.

    Although we do not know how to calculate $\chi_1$ and $\chi_2$ from first
principles, we do anticipate that [3]
\be
|\chi(B\to PP)|<|\chi(D\to PP)|<|\chi(D\to VP)|,
\en
based on the reason that nonperturbative soft gluon effects become more
important when the final-state particles move slower, allowing more time for
significant final-state interactions after hadronization. As a consequence,
it is obvious that $a_{1,2}$ are in general not universal and that
the rule of discarding $1/N_c$ terms [5], which works
empirically well in $D\to\bar{K}\pi$ decay, cannot be safely extrapolated to
$B\to D\pi$ decay as $|\chi(B\to D\pi)|$ is expected to be much smaller than
$|\chi(D\to\bar{K}\pi)\sim -{1\over 3}|$ (the c.m. momentum in $D\to\bar{K}
\pi$ being 861 MeV, to be compared with 2307 MeV in $B\to D\pi$)
and hence a large cancellation between
$1/N_c$ and $\chi(B\to D\pi)$ is not expected to happen. The recent CLEO
observation [6] that the rule of discarding $1/N_c$ terms is not operative
in $B\to D(D^*)\pi(\rho)$ decays is therefore not stunning. Only the fact
that $\chi(B\to D\pi)$ is positive turns out to be striking.

  Unlike the $PP$ or $VP$ case, it is not pertinent to define $\chi_{1,2}$
for $M\to VV$ decay as its general amplitude consists of three independent
Lorentz scalars:
\be
A[M(p)\to V_1(\ep_1,p_1)V_2(\ep_2,p_2)]\propto \ep^*_\mu(\lambda_1)
\ep^*_\nu(\lambda_2)(\hat{A}_1g^{\mu\nu}+\hat{A}_2p^\mu p^\nu+i
\hat{V}\epsilon^{\mu\nu\alpha\beta}p_{1\alpha}p_{2\beta}),
\en
where $\hat{A}_1,~\hat{A}_2,~\hat{V}$ are related to the form factors
$A_1,~A_2$ and $V$ respectively. Since {\it a priori} there is no reason to
expect that nonfactorizable terms weight in the same way to $S$-, $P$-
and $D$-waves, namely
$A_1^{nf}/A_1=A_2^{nf}/A_2=V^{nf}/V$, we thus cannot define
$\chi_1$ and $\chi_2$. Consequently, it is in general not possible to define
an effective $a_1$ or $a_2$ for $M\to VV$ decays once nonfactorizable
effects are taken into account [7]. In the factorization approach, the
fraction of polarization, say $\Gamma_L/\Gamma$ ($L$: longitudinal
polarization) in $B\to\psi K^*$ decay, is independent of the parameter $a_2$.
As a result, if an effective $a_2$ can be defined for $B\to \psi K^*$, it
will lead to the conclusion that nonfactorizable terms cannot
affect the factorization prediction of $\Gamma_L/\Gamma$ at all. It was
realized recently that all the known models in the literature in conjunction
with the factorization hypothesis fail to reproduce the data of $\Gamma_L/
\Gamma$ or the production ratio $\Gamma(B\to\psi K^*)/\Gamma(B\to\psi K)$
or both [8,9].
Evidently, if we wish to utilize nonfactorizable effects to
resolve the puzzle with $\Gamma_L/\Gamma$, a key
ingredient will be the nonexistence of an effective $a_2$ for $B\to\psi K^*$.

   In short, there are two places where the factorization hypothesis can be
unambiguiously tested: (i) To extract the parameters $a_1$ and $a_2$ from
the experimental measurements of $M\to PP,~VP$ to see if they are process
independent. (ii) To measure the fraction of longitudinal polarization in $M\to
VV$ decay and compare with the factorization prediction. Any failure of them
will indicate a breakdown of factorization.

The purpose of the present paper is threefold. (i) The parameters $\chi_1$ and
$\chi_2$ have been extracted in Ref.[3] (see also [10]). Here we wish to
update the values of $\chi_{1,2}$ using the $q^2$ dependence of form factors
suggested by QCD-sum-rule calculations and other theoretical arguments.
(ii) It was recently advocated by Kamal and Sandra [7] that the assumption
that in $B\to \psi K^*$ decay the nonfactorizable amplitude contributes
only to $S$-wave final states, namely $A_1^{nf}\neq 0,~A_2^{nf}=V^{nf}=0$,
will lead to a satisfactory explanation of the data of $\Gamma(B\to \psi K^*)/
\Gamma(B\to \psi K)$ and $\Gamma_L/\Gamma$. We would like to show that this
very assumption is also essential for understanding the ratio ${\cal B}(B^-
\to D^{*0}\rho^-)/{\cal B}(\bar{B}^0\to D^{*+}\rho^-)$, which cannot be
explained satisfactorily in previous work. (iii) Contrary to the $B$ meson
case, we will demonstrate that the assumption of $S$-wave dominated
nonfactorizable terms does not work in $D\to VV$ decay.

\vskip 0.30 cm
\noindent{\bf 2.~~Nonfactorizable contributions in $D,~B\to PP,~VP$ decays}~~

   Because of the presence of final-state interactions (FSI) and the
nonspectator contributions ($W$-exchange and $W$-annihilation), it is
generally not possible to extract the nonfactorization parameters $\chi_{1,2}$
except for a very few channels. Though color-suppressed decays, for example,
$D^0\to\bar{K}^0(\bar{K}^{*0})\pi^0(\rho^0)$ are conventionally classified as
Class II modes [11], color-flavored decay $D^0\to K^-\pi^+$ will bring some
important contribution to $D^0\to\bar{K}^0\pi^0$ via FSI. This together
with the small but not negligible $W$-exchange amplitude renders the
determination of
$a_2$ from $D^0\to\bar{K}^0\pi^0$ impossible. Therefore, in order to determine
$a_1$ and especially $a_2$ we should focus on the exotic channels e.g.
$D^+\to\bar{K}^0\pi^+,~\pi^+\pi^0$, and the decay modes with one single isospin
component, e.g. $D^+\to\pi^+\phi,~D_s^+\to\pi^+\phi$, where nonspectator
contributions are absent and FSI are presumably negligible.

   We next write down the relations between $\chi_{1,2}$ and form factors
\be
 \chi_1(D\to\bar{K}\pi) = {F_0^{nf}(m_\pi^2)\over F_0^{DK}(m_\pi^2)},&& ~~~~
\chi_2(D\to\bar{K}\pi) = {F_0^{nf}(m_K^2)\over F_0^{D\pi}(m_K^2)}, \non \\
\chi_1(D\to\bar{K}^*\pi) = {A_0^{nf}(m_\pi^2)\over A_0^{DK^*}(m_\pi^2)},&& ~~~~
\chi_2(D\to\bar{K}^*\pi) = {F_1^{nf}(m_{K^*}^2)\over F_1^{D\pi}(m_{K^*}^2)},
 \non \\
\chi_1(D\to\bar{K}\rho)= {F_1^{nf}(m_\rho^2)\over F_1^{DK}(m_\rho^2)},&&~~~~
\chi_2(D\to\bar{K}\rho) = {A_0^{nf}(m_{K}^2)\over A_0^{D\rho}(m_{K}^2)},
  \\
\chi_1(D^+_s\to\phi\pi^+) = {A_0^{nf}(m_\pi^2)\over A_0^{D_s\phi}(m_\pi
^2)},&&~~~~\chi_2(D^+\to\phi\pi^+) = {F_1^{nf}(m_\phi^2)\over F_1^{D\pi}
(m^2_\phi)}.   \non
\en
It is clear that only the three form factors $F_0,~F_1$ and $A_0$ entering
into the decay amplitudes of $M\to PP,~VP$. A consideration of the heavy
quark limit behavior of the form factors suggests that the $q^2$ dependence
of $F_1~(A_2)$ is different from that of $F_0$ ($A_0$ and $A_1$) by an
additional pole factor [12]. Indeed, QCD-sum-rule calculations have implied
a monopole behavior for $F_1(q^2)$ [13-16] and an approximately constant
$F_0$ [15]. With a dipole form factor $A_2$, as shown by a recent QCD-sum-rule
analysis [16], we will thus assume a monopole behavior for $A_0$.

    Unlike the decays $D^+\to\pi^+\phi,~D_s^+\to\pi^+\phi$ which are described
by a single quark diagram, we cannot extract
$\chi_{1,2}$ from the data of $D^+\to\bar{K}^0\pi^+,
{}~\bar{K}^0\rho^+,~\bar{K}^{*0}\pi^+$ alone without providing further
information. For example, the decay amplitude of $D^+\to\bar{K}^0\pi^+$ reads
\be
A(D^+\to\bar{K}^0\pi^+)=\,{G_F\over\sqrt{2}}V_{cs}^*V_{ud}[a_1(m_D^2-m_K^2)
f_\pi F_0^{DK}(m_\pi^2)+a_2(m_D^2-m_\pi^2)f_KF_0^{D\pi}(m_K^2)],
\en
which consists of external $W$-emission and internal $W$-emission amplitudes.
We will therefore make a plausible assumption that $\chi_1\sim\chi_2$ so that
$\chi(D\to\bar{K}\pi)$ can be determined from the measured rate of $D^+\to
\bar{K}^0\pi^+$. Since the extraction procedure is already elucidated in
Ref.[3], here we will simply present the results (only the central values
being quoted) followed by several remarks
\be
\chi_2(D\to\bar{K}\pi) &\simeq & -0.36\,,   \non \\
\chi_2(D\to\bar{K}^*\pi) &\simeq & -0.61\,,   \\
\chi_2(D^+\to\phi\pi^+) &\simeq & -0.44\,,   \non
\en
where we have used the following quantities:
\be
&& c_1(m_c)=1.26,~~~~~c_2(m_c)=-0.51,   \non \\
&& f_\pi=132\,{\rm MeV},~~~f_K=160\,{\rm MeV},~~~f_{K^*}=220\,{\rm MeV},~~~
f_\phi=237\,{\rm MeV},   \non \\
&& F_0^{DK}(0)=F_1^{DK}(0)=0.77\pm 0.04~[17],~~~~F_0^{D\pi}(0)=F_1^{D\pi}(0)
=0.83~[18],    \\
&& A_1^{DK^*}(0)=0.61\pm 0.05,~~~~A_2^{DK^*}(0)=0.45\pm 0.09~[17],~\Rightarrow
A_0^{DK^*}(0)=0.70,    \non
\en
and the Particle Data Group [19] for the decay rates of various decay modes.

   Several remarks are in order. (i) As pointed out by Soares [10], the
solutions for $\chi$ are not uniquely determined. For example, the other
possible solution for $\chi_2(D\to\bar{K}\pi)$ is $-1.18\,$. To remove
the ambiguities, we
have assumed that nonfactorizable corrections are small compared to the
factorizable ones. (ii) Assuming $A_0^{D\rho}(0)=A_0^{DK^*}(0)$, we find from
the decay $D^+\to\bar{K}^0\rho^+$ that $\chi(D\to\bar{K}\rho)\approx -1.5$,
wich is unreasonably too large. We do not know how to resolve this problem
except for noting that thus far there is only one measurement of this decay
mode [20]. (iii) To determine $\chi_1(D_s^+\to\phi\pi^+)$
requires a better knowledge of the form factor $A_0^{D_s\phi}$ and the
branching ratio of $D_s^+\to\phi\pi^+$. Unfortunatly, a direct measurement of
them is still not available. Assuming $A_0^{D_s\phi}(0)\approx A_0^{DK^*}(0)$
and ${\cal B}(D_s^+\to\phi\pi^+)=(3.5\pm 0.4)\%$ [19], we get $\chi_1(D_s^+
\to\phi\pi^+)\approx -0.60\,$. So in general nonfactorizable terms
are process
or class dependent, and satisfy the relation $|\chi(D\to PP)|<|\chi(D\to
VP)|$ as expected. (iv) Since $\chi_2(D\to \bar{K}\pi)$ is close to $-{1\over
3}$, it is evident that a large cancellation between $1/N_c$ and $\chi_2(D\to
\bar{K}\pi)$ occurs. This is the dynamic reason why the large-$N_c$ approach
operates well for $D\to\bar{K}\pi$ decay. However, this is no longer the
case for $D\to VP$ decays. The predicted branching ratios in $1/N_c$
expansion are
\be
{\cal B}(D^+\to\bar{K}^{*0}\pi^+)=0.3\%,&&~~~~{\cal B}(D^+\to\bar{K}^0
\rho^+)= 16\%,   \non \\
{\cal B}(D^+\to\bar{K}^{*0}\rho^+)= 17\%,&&~~~~{\cal B}(D^+\to
\phi\pi^+)= 0.4\%,
\en
to be compared with data [19]
\be
{\cal B}(D^+\to\bar{K}^{*0}\pi^+)_{\rm expt}= (2.2\pm 0.4)\%,&&~~~~{\cal B}
(D^+\to\bar{K}^0\rho^+)_{\rm expt}=(6.6\pm 2.5)\%,  \non \\
{\cal B}(D^+\to\bar{K}^{*0}\rho^+)_{\rm expt}= (4.8\pm 1.8)\%,&&~~~~{\cal
B}(D^+\to\phi\pi^+)_{\rm expt}= (0.67\pm 0.08)\%.
\en
Consider the decay $D^+\to\bar{K}^{*0}\pi^+$ as an example. Its amplitude is
given by
\be
A(D^+\to\bar{K}^{*0}\pi^+)=\sqrt{2}G_FV_{cs}^*V_{ud}[a_1f_\pi m_{K^*}A_0^{DK^*}
(m_\pi^2)+a_2f_{K^*}m_{K^*}F_1^{D\pi}(m^2_{K^*})].
\en
Since the interference is destructive and $f_{K^*}F_1^{D\pi}>f_\pi A_0^{DK^*}$,
a large $|a_2|$ is needed in order to enhance the branching ratio of
$D^+\to\bar{K}^{*0}\pi^+$ from 0.3\% to 2.2\%. (Note that $a_1$ is relatively
insensitive
to the nonfactorizable effects.) This in turn implies a negative (${1\over
N_c}+\chi_2)$ and hence $\chi_2(D\to\bar{K}^*\pi)<-{1\over 3}$.
Therefore, we are led to conclude that the leading $1/N_c$ expansion
cannot be a universal approach for the nonleptonic weak decays of the meson.
However, the fact that substantial nonfactorizable effects which contribute
destructively with the subleading $1/N_c$ factorizable contributions are
required to accommodate the data of charm decay means that, as far as
charm decays are concerned, the large-$N_c$ approach greatly improves the
naive factorization method in which $\chi_{1,2}=0$; the former approach
amounts to having a universal nonfactorizable term $\chi_{1,2}=-1/N_c$.

  We next turn to $B\to D(D^*)\pi(\rho)$ decays. Though both nonspectator and
FSI effects are known to be important in charm decays, it is generally
believed that they do not play a significant role in bottom decays as
the decay particles are moving fast, not allowing adequate time for FSI.
This gives the enormous advantage that it is conceivable to determine $a_1$
and $a_2$ separately from $B\to D(D^*)\pi(\rho)$ decays. Using the
heavy-flavor-symmetry approach for heavy-light form factors and assuming
a monopole extrapolation for $F_1,~A_0,~A_1$, a dipole behavior for $A_2,~V$,
and an approximately constant $F_0$, as suggested by QCD-sum-rule calculations
and some theoretical arguments [21], we found from the CLEO data that [21]
\footnote{Contrary to the charmed meson case, the variation of $a_{1,2}$
from $B\to D\pi$ to $D^*\pi$ and $D\rho$ decays is negligible (see Table IV of
[21]).}
\be
a_1(B\to D^{(*)}\pi(\rho)) &=& 1.01\pm 0.06\,,  \non \\
a_2(B\to D^{(*)}\pi(\rho)) &=& 0.23\pm 0.06\,.
\en
Taking $c_1(m_b)=1.11$ and $c_2(m_b)=-0.26$ leads to
\be
\chi_1(B\to D^{(*)}\pi(\rho))\simeq 0.05\,,~~~~\chi_2(B\to D^{(*)}\pi(\rho))
\simeq 0.11\,.
\en
Since (${1\over N_c}+\chi_{1,2})=(a_{1,2}-c_{1,2})/c_{2,1}$ and
$|c_2|<<|c_1|$, it
is clear that the determination of $\chi_1$ is far more uncertain than
$\chi_2$: it is very sensitive to the values of $a_1,~c_1$ and $c_2$. We see
from (18) that nonfactorizable effects become less important in $B$ decays,
as what expected [see (8)]. However, a positive $\chi_2(B\to D(D^*)\pi(
\rho))$, which
is necessary to explain the constructive interference in $B^-\to D^0(D^{*0})
\pi^-(\rho^-)$ decays, appears to be rather striking. A recent light cone
QCD-sum-rule calculation [22] following the framework outlined in [23] fails
to reproduce a positive $\chi_2(B\to D\pi)$. This tantalizing issue should
be resolved in the near future.

   For $B\to\psi K$ decays, we found [21]
\be
\left|a_2(B^-\to\psi K^-)\right|=\,0.235\pm 0.018\,,~~~~
\left|a_2(B^0\to\psi K^0)\right|=\,0.192\pm 0.032\,.
\en
The combined value is
\be
a_2(B\to\psi K)=\,0.225\pm 0.016\,,
\en
where its sign should be positive, as we have argued in
[21]. (It was advocated by Soares [10] that an analysis of the contribution
of $B\to\psi K$ to the decay $B\to K\ell^+\ell^-$ can be used to remove the
sign ambiguity of $a_2$.) It follows that
\be
\chi_2(B\to\psi K)={F_1^{nf}(m_\psi^2)\over F_1^{BK}(m_\psi^2)}\simeq 0.10\,,
\en
which is in accordance with $\chi_2(B\to D^{(*)}\pi(\rho))$.

   Finally, it is very interesting to note that, in contrast to charm decays,
the large-$N_c$ approach is even worse than the naive factorization method
in describing $B\to D(D^*)\pi(\rho)$ decays as
$\chi_2(B\to D^{(*)}\pi(\rho))$ is small but positive.

\vskip 0.30cm
\noindent{\bf 3.~~Nonfactorizable contributions in $B\to \psi K^*,~D^*\rho$
decays}

   As stressed in the Introduction, in general one cannot define $\chi_{1,2}$
and hence an effective $a_{1,2}$ for $M\to VV$ decays unless the
nonfactorizable terms weight in the same manner in all three partial waves.
It was pointed out recently that there are two
experimental data, namely the production ratio $R\equiv \Gamma(B\to\psi K^*)/
\Gamma(B\to\psi K)$ and the fraction of longitudinal polarization $\Gamma_L/
\Gamma$ in $B\to\psi K^*$, which cannot be accounted for simultaneously by all
commonly used models within the framework of factorization [8,9]. The
experimental results are
\be
R=\,1.74\pm 0.39~[6],~~~~{\Gamma_L\over\Gamma}=\,0.78\pm 0.07\,,
\en
where the latter is the combined average of the three measurements:
\be
\left({\Gamma_L\over\Gamma}\right)_{B\to\psi K^*}=\cases{ 0.97\pm 0.16\pm
0.15, & ARGUS~[24]; \cr 0.80\pm 0.08\pm
0.05, & CLEO~[6]; \cr 0.66\pm 0.10^{+0.08}_{-0.10},& CDF~[25].  \cr}
\en

Irrespective of the production ratio $R$, all the existing models fail to
produce a large longitudinal polarization fraction [8,9]. This strongly
implies that the puzzle with $\Gamma_L/\Gamma$ can only be resolved by
appealing to nonfactorizable effects.
\footnote{An interesting observation was made recently in [26] that the
factorization assumption in $B\to\psi
K(K^*)$ is not ruled out and the data can be accommodated by the
heavy-flavor-symmetry approach for heavy-light form factors provided that
the $A_1(q^2)$ form factor is frankly decreasing. To our knowledge, a
decreasing $A_1$ with $q^2$ is ruled out by several recent QCD-sum-rule
analyses (see e.g. [16]). Using the same approach for heavy-light form
factors but the $q^2$ dependence of form factors given in [21], we found
that $R=1.84$ and $\Gamma_L/\Gamma
=0.56$ [21]. Evidently, the factorization approach is still difficult to
explain the observed large polarization fraction.}
However, if the relation $A_1^{nf}/A_1=A_2^{nf}/A_2=V^{nf}/V$ holds, then an
effective $a_2$ can be defined for $B\to\psi K^*$ and the prediction of
$\Gamma_L/\Gamma$ will be the same as that in the factorization approach as
the polarization fraction is independent of $a_2$. Consequently,
nonfactorizable terms should contribute differently to $S$-, $P$- and
$D$-wave amplitudes if we wish to explain the observed $\Gamma_L/\Gamma$.

   The large longitudinal polarization fraction observed by ARGUS and CLEO
suggests that the decay $B\to\psi K^*$ is almost all $S$-wave. To see this, we
write down the $B\to\psi K^*$ amplitude
\be
A[B(p)\to\psi(p_1)K^*(p_2)] &=& {G_F\over\sqrt{2}}V_{cs}^*V_{ud}\left(c_2+
{c_1\over 3}\right)f_\psi m_\psi \ep^*_\mu(\psi)
\ep^*_\nu(K^*)[\hat{A}_1g^{\mu\nu}+\hat{A}_2p^\mu p^\nu   \non \\
&+& i\hat{V}\epsilon^{\mu\nu\alpha\beta}p_{1\alpha}p_{2\beta}],
\en
with
\be
\hat{A}_1 &=& (m_B+m_{K^*})A_1^{BK^*}(m_\psi^2)\left[1+\kappa{A_1^{nf}(m_
\psi^2)\over A_1^{BK^*}(m_\psi^2)}\right],  \non \\
\hat{A}_2 &=& -{2\over (m_B+m_{K^*})}A_2^{BK^*}(m_\psi^2)\left[1+\kappa{A_2^
{nf}(m_\psi^2)\over A_2^{BK^*}(m_\psi^2)}\right],   \\
\hat{V} &=& -{2\over (m_B+m_{K^*})}V^{BK^*}(m_\psi^2)\left[1+\kappa{V^
{nf}(m_\psi^2)\over V^{BK^*}(m_\psi^2)}\right], \non
\en
and $\kappa=c_1/(c_2+{1\over 3}c_1)$. It is easily seen that we  will have an
effective $a_2=c_2+c_1({1\over 3}+\chi_2)$ if the nonfactorizable
terms happen to satisfy the relation $A_1^{nf}/A_1=A_2^{nf}/A_2=V^{nf}/V=
\chi_2$. The decay rate of this mode is of the form
\be
\Gamma(B\to\psi K^*)\propto \,(a-b\tilde{x})^2+2(1+c^2\tilde{y}^2),
\en
where
\be
a={m^2_B-m^2_\psi-m^2_{K^*}\over 2m_\psi m_{K^*}},&&~~b= {2m_B^2p_c^2\over
m_\psi m_{K^*}(m_B+m_{K^*})^2},~~~c={2m_Bp_c\over (m_B+m_{K^*})^2},   \non\\
\tilde{x}={A_2^{BK^*}(m_\psi^2)+\kappa A_2^{nf}(m_\psi^2)\over A_1^{BK^*}
(m_\psi^2)+\kappa A_1^{nf}(m_\psi^2)},&&~~\tilde{y}={V^{BK^*}(m_\psi^2)+\kappa
V^{nf}(m_\psi^2)\over A_1^{BK^*}(m_\psi^2)+\kappa A_1^{nf}(m_\psi^2)},
\en
with $p_c$ being the c.m. momentum. The longitudinal polarization fraction
is then given by
\be
{\Gamma_L\over \Gamma}={(a-b\tilde{x})^2\over (a-b\tilde{x})^2+2(1+c^2\tilde
{y}^2)}.
\en
If the decay is an almost $S$-wave, one will have $\Gamma_L/\Gamma\sim
a^2/(a^2+2)=0.83\,$. Since $\kappa>>1$, $\tilde{x}$ ($D$-wave) and
$\tilde{y}$ ($P$-wave) can be suppressed by assuming that,
as first postulated in [7], in $B\to\psi K^*$ decay the nonfactorizable
amplitude contributes only to $S$-wave final states; that is,
\footnote{ A different approach for nonfactorizable effects adopted in Ref.[27]
amounts to $A_1^{nf}=A_2^{nf}=0$ and $V^{nf}\neq 0$. It follows from Eq.(28)
that
\be
{\Gamma_L\over \Gamma}={(a-b{x})^2\over (a-b{x})^2+2(1+c^2\bar{y}^2)}, \non
\en
with $\bar{y}=(V^{BK^*}(m_\psi^2)+\kappa V^{nf}(m_\psi^2))/A_1^{BK^*}(m_
\psi^2)$ and $x$ being defined in (31). It is clear that in order to get a
large longitudinal polarization fraction one needs a {\it negative} $V^{nf}/V$
! Using the
numerical values $a=3.164,~b=1.304,~x=0.89$, we find $(\Gamma_L/\Gamma)_{\rm
max}=0.67$. The prediction $\Gamma_L/\Gamma=0.65$ given by [27] is one standard
deviation from experiment (22).}
\be
A^{nf}_1\neq 0,~~~~A_2^{nf}=V^{nf}=0.
\en
The rational for this assumption is given in [7].

With the assumption (29), the branching ratio followed from (24) is
\be
{\cal B}(B\to\psi K^*)=0.0288\left|\left(c_2+{c_1\over 3}\right)A_1^{BK^*}
(m_\psi^2)\right|^2[(a\xi-bx)^2+2(\xi^2+c^2y^2)]
\en
with
\be
x={A_2^{BK^*}(m_\psi^2)\over A_1^{BK^*}(m_\psi^2)},~~~y={V^{BK^*}(m_\psi^2)
\over A_1^{BK^*}(m_\psi^2)},~~~\xi=1+\kappa{A_1^{nf}(m_\psi^2)\over
A_1^{BK^*}(m_\psi^2)},
\en
where uses of $|V_{cb}|=0.040$ and $\tau(B)=1.52\times 10^{-12}s$ have been
made. It follows that
\be
{\Gamma_L\over \Gamma}=\,{(a\xi-bx)^2\over (a\xi-bx)^2+2(1+c^2y^2)}.
\en
We use the measured branching ratio ${\cal B}(B\to\psi K^*)=(0.172
\pm 0.030)\%$ [6] to determine the ratio $A_1^{nf}(m_\psi^2)/A_1^{BK^*}(m_\psi
^2)$, which is found to be
\be
{A_1^{nf}(m_\psi^2)\over A_1^{BK^*}(m_\psi^2)}\simeq 0.08\,,
\en
which we have used $A_1^{BK^*}(m_\psi^2)=0.41,~A_2^{BK^*}(m_\psi^2)=0.36,~
V^{BK^*}(m_\psi^2)=0.72$ [21] and discarded the other possible solution
$A_1^{nf}/A_1^{BK^*}=-0.22$ for its ``wrong'' sign, recalling that
$F_1^{nf}/F_1^{BK}$ is positive [cf. Eq.(21)].
The predicted longitudinal polarization fraction is $\Gamma_L/
\Gamma=0.73$, which is in accordance with experiment.

    The assumption of negligible nonfactorizable contributions to $P$- and
$D$-waves also turns out to be essential for understanding the decay rate of
$B^-\to D^{*0}\rho^-$ or the ratio $R_4\equiv{\cal B}(B^-\to D^{*0}\rho^-)/
{\cal B}(\bar{B}^0\to D^{*+}\rho^-)$. The issue arises as follows. In Ref.[21]
we have determined $a_1$ and $a_2$ from $B\to D\pi,~D^*\pi,~D\rho$ decays
and obtained a consistent ratio $a_2/a_1$ from channel to channel:
$0.24\pm 0.10,~0.24\pm 0.14,~0.21\pm 0.08$ (see Table IV of [21]).
Assuming factorization, we got $a_2/a_1=0.34\pm 0.13$ from $B\to D^*\rho$
decay, which deviates somewhat from above values. In the presence of
$S$-wave dominated nonfactorizable contributions, it is no longer possible
to define an effective $a_1$ and $a_2$ for $B\to D^*\rho$ decay. Therefore, the
quantities to be compared with are $A_1^{nf}/A_1$ in $B\to D^*\rho$ decay and
$\chi_2$ in $B\to D\pi,~D^*\pi,~D\rho$. A straightforward calculation yields
(see [21] for the factorizable case)
\be
R_4={\tau(B^-)\over\tau(B^0)}\left(1+2{\eta}{{H}_1\over {H}}
+{\eta}^2{{H}_2\over {H}}\right),
\en
with
\be
H &=& (\hat{a}\hat{\xi}-\hat{b}\hat{x})^2+2(\hat{\xi}^2+\hat{c}^2\hat{y}^2),
\non \\
{H}_1 &=& (\hat{a}\hat{\xi}-\hat{b}\hat{x})(\hat{a}\hat{\xi}'-\hat{b}'\hat{x}'
)+2(\hat{\xi}\hat{\xi}'+\hat{c}\hat{c}'\hat{y}\hat{y}'),   \non \\
H_2 &=& (\hat{a}\hat{\xi}'-\hat{b}'\hat{x}')^2+2(\hat{\xi}'^2+\hat{c}'^2
\hat{y}'^2),    \non \\
\eta &=& {m_{D^*}(m_B+m_\rho)\over m_\rho(m_B+m_{D^*})}\,{f_{D^*}\over
f_\rho}\,{A_1^{B\rho}(m^2_{D^*})\over A_1^{BD^*}(m^2_\rho)}\,{c_2+{1\over 3}
c_1\over c_1+{1\over 3}c_2},    \\
\hat{\xi} &=& 1+{c_2\over c_1+{1\over 3}c_2}\,{A_1^{nf}(m_{\rho}^2)\over A_1^
{BD^*}(m_{\rho}^2)},   \non \\
\hat{\xi}' &=& 1+{c_1\over c_2+{1\over 3}c_1}\,{A_1^{nf}(m_{D^*}^2)\over A_1^
{B\rho}(m_{D^*}^2)},  \non
\en
where $\hat{a},~\hat{b},~\hat{c}$ are obtained from $a,~b,~c$ respectively
in (27), $\hat{x},~\hat{y}$ from $x,~y$ in (31) by replacing $\psi\to D^*,
{}~K^*\to \rho$,
and $\hat{b}',~\hat{c}',~\hat{x}',~\hat{y}'$ are obtained from $\hat{b},~
\hat{c},~\hat{x},~\hat{y}$ respectively by replacing
$D^*\leftrightarrow \rho$; for instance $\hat{x}'=A_2^{B\rho}(m^2_{D^*})/
A_1^{B\rho}(m_{D^*}^2)$. Assuming $A_1^{nf}/A_1^{BD^*}\sim A_1^{nf}/A_1^{B
\rho}$ and fitting (34) to the experimental value $R_4=(1.68\pm 0.35)\%$ [6],
we get
\be
{A_1^{nf}(m^2_{D^*})\over A_1^{B\rho}(m_{D^*}^2)}\sim
{A_1^{nf}(m^2_{\rho})\over A_1^{BD^*}(m_{\rho}^2)}\simeq 0.12\,.
\en
We see that the $S$-wave dominated nonfactorizable effect in $B\to\psi K^*$ and
$B\to D^*\rho$ decays is of order 10\%, consistent with $\chi_2(B\to\psi K)$
and $\chi_2(B\to D(D^*)\pi(\rho))$.

\vskip 0.30cm
\noindent{\bf 4.~~Nonfactorizable contributions in $D\to\bar{K}^*\rho$ decay}

   We have shown in the previous section that $S$-wave dominated
nonfactorizable terms are needed to explain the large longitudinal
polarization fraction observed in $B\to\psi K^*$ and the ratio ${\cal B}
(B^-\to D^{*0}\rho^-)/{\cal B}(\bar{B}^0\to D^{*+}\rho^-)$. However, we
shall see in this section that the assumption (29) is no longer applicable to
$D\to \bar{K}^*\rho$ decay. An experimental measurement of $D^+\to \bar{K}
^{*0}\rho^+$ and $D^0\to\bar{K}^{*0}\rho^0$ by Mark III [28] shows that
(i) the decay $D^+\to \bar{K}^{*0}\rho^+$
is a mixture of longitudinal and transverse polarization
consistent with a pure $S$-wave amplitude,
\footnote{The other measurement by E691 [29] disagrees severely with Mark
III on the branching ratio
\be
{\cal B}(D^+\to\bar{K}^{*0}\rho^+)=\cases{(4.8\pm 1.2\pm 1.4)\%,
&Mark~III~[28];
\cr  (2.3\pm 1.2\pm 0.9)\%, &E691~[29].    \cr}
\en
Recall that model calculations tend to give a very large branching ratio
of 17\% [see Eq.(14)].}
and (ii) $D^0\to\bar{K}^{*0}\rho^0$
is almost all transverse, requiring a cancellation between the longitudinal
$S$-wave and $D$-wave.

    We first consider the decay $D^+\to\bar{K}^{*0}\rho^+$, whose amplitude is
given by
\be
A(D^+(p)\to\bar{K}^{*0}(p_1)\rho^+(p_2))={G_F\over\sqrt{2}}V_{cs}^*V_{ud}
\ep^*_\mu(K^*)\ep^*_\nu(\rho)[\tilde{A}_1g^{\mu\nu}+\tilde{A}_2p^\mu p^\nu+i
\tilde{V}\epsilon^{\mu\nu\alpha\beta}p_{1\alpha}p_{2\beta}],
\en
where
\be
\tilde{A}_1 &=& \left(c_1+{c_2\over 3}\right)f_\rho m_\rho(m_D+m_{K^*})\left
(1+{c_2\over
c_1+{1\over 3}c_2}{A_1^{nf}(m_\rho^2)\over A_1^{DK^*}(m_\rho^2)}\right)
A_1^{DK^*}(m_\rho^2)   \non \\
&+& \left(c_2+{c_1\over 3}\right)f_{K^*} m_{K^*}(m_D+m_\rho)\left(1+{c_1\over
c_2+{1\over 3}c_1}{A_1^{nf}(m_{K^*}^2)\over A_1^{D\rho}(m_{K^*}^2)}\right)
A_1^{D\rho}(m_{K^*}^2),
\en
and $\tilde{A}_2$ ($\tilde{V}$) is obtained from $\tilde{A}_1$ with the
replacements $A_1\to A_2$ ($A_1\to V$),  $(m_D+m_{K^*})\to -2/
(m_D+m_{K^*})$ and $(m_D+m_\rho)\to -2/(m_D+m_\rho)$.
Since $A_1^{nf}/A_1^{DK^*}$ and $A_1^{nf}/A_1^{D\rho}$ are expected to be
negative [see Eq.(12)], it is obvious that if nonfactorizable terms
are dominated by the $S$-wave, it will imply a more severe destructive
interference in the $S$-wave amplitude than in $P$- and $D$-wave amplitudes,
in contradiction to the observation
that this decay is almost all $S$-wave. The branching ratio is calculated
to be
\be
{\cal B}(D^+\to\bar{K}^{*0}\rho^+)=\,0.10\left|\left(c_1+{1\over 3}c_2\right)
A_1^{DK^*}(m_\rho^2)\right|^2(H'+2\eta'H'_1+\eta'^2H'_2),
\en
with the expressions of $\eta',~H',~H_{1,2}'$ analogous to $\eta,~H,~H_{1,2}$
in (35). A fit of (40) to the Mark III data for the branching ratio (37)
gives rise to (assuming $A_1^{nf}/A_1^{DK^*}\sim A_1^{nf}/A_1^{D\rho}$)
\be
{A_1^{nf}(m^2_{\rho})\over A_1^{DK^*}(m_{\rho}^2)}\sim
{A_1^{nf}(m^2_{K^*})\over A_1^{D\rho}(m_{K^*}^2)}\approx -0.98\,,
\en
which is uncomfortably too large.
\footnote{A fit to the E691 measurement (37) for the branching ratio yields an
even larger value: $A_1^{nf}/A_1^{DK^*}\sim A_1^{nf}/A_1^{D\rho}\approx
-1.41\,$.}
Moreover, the $P$-wave branching ratio is
predicted to be $2.0\times 10^{-2}$, in disagreement with experiment [28]
\be
{\cal B}(D^+\to\bar{K}^{*0}\rho^+)_{P{\rm -wave}}<0.5\times 10^{-2}.
\en
It thus appears to us that an almost $S$-wave $D^+\to\bar{K}^{*0}\rho^+$
implies that
\be
\left|{A_2^{nf}\over A_2^{DK^*(\rho)}}\right|,~~~\left|{V^{nf}\over V^{DK^*(
\rho)}}\right| {\ \lower-1.2pt\vbox{\hbox{\rlap{$>$}\lower5pt\vbox{\hbox{$
\sim$}}}}\ }\left|{A_1^{nf}\over A_1^{DK^*(\rho)}}\right|.
\en
Taking $A_1^{nf}/A_1=A_2^{nf}/A_2=V^{nf}/V=\chi(D\to \bar{K}^*\rho)$ as an
illustration, we obtain
\be
\chi(D\to\bar{K}^*\rho)\approx -0.65\,
\en
and ${\cal B}(D^+\to\bar{K}^{*0}\rho^+)_{P{\rm -wave}}=2.0\times 10^{-3}$,
which are certainly more plausible than before.

  Another indication for the failure of the $S$-wave dominated hypothesis for
nonfactorizable effects comes from the decay $D^0\to\bar{K}^{*0}\rho^0$,
where $\bar{K}^{*0}$ and $\rho^0$ are completely transversely polarized,
implying a large $D$-wave which is compensated by the longitudinal $S$-wave.
Recall that the factorizable $D\to VV$ amplitudes have the sailent feature :
\be
|S{\rm -wave~amplitude}|>|P{\rm -wave~amplitude}|>|D{\rm -wave~amplitude}|.
\en
Since the color-suppressed $D$-wave amplitude of $D^0\to\bar{K}^{*0}\rho^0$
is proportional to $[1+c_1/(c_2+{1\over 3}c_1)(A_2^{nf}/A_2^{D\rho})]$,
a large $D$-wave thus indicates a negative $A_2^{nf}/A_2$ and
\be
\left|{A_1^{nf}\over A_1}\right|<<\left|{A_2^{nf}\over A_2}\right|,~~~~{\rm or}
{}~~{A_1^{nf}\over A_1}\approx 0,~~{A_2^{nf}\over A_2}\neq 0.
\en
Therefore, we see that nonfactorizable terms in charm decay are consistently
to be negative [cf. Eqs.(12) and (44)].
Unfortunately, at this point we cannot make a further quantitative analysis
due to unknown final-state interactions and $W$-exchange contributions.
A measurement of helicities in $D^0\to\bar{K}^{*0}\rho^0,~D^+\to\phi\rho^+$
will be greatly helpful to pin down the issue. In particular, the color- and
Cabibbo-suppressed
mode $D^+\to\phi\rho^+$ is very ideal for this purpose since it
is not subject to FSI and nonspectator effects. A polarization measurement
in this decay is thus strongly urged (though difficult) in order to test if
$A_2^{nf}/A_2$ plays a more essential role than $A_1^{nf}/A_1$ in charm decay.

\vskip 0.30 cm
\noindent{\bf 5.~~Discussion and conclusion}

    The factorization assumption for hadronic weak decays of mesons
can be tested on two different grounds: (i) to extract the effective parameters
$a_1$ and especially $a_2$ from $M\to PP,~VP$ decays to see if they are
process independent, and (ii) to measure helicities in $M\to VV$
decay. Using the $q^2$ dependence of form factors suggested by QCD-sum-rule
calculations and by some theoretical arguments, we have updated our
previous work.
It is found that $a_2$ is evidently not universal in charm decay. The
parameter $\chi_{2}$, which measures the nonfactorizable soft-gluon effect on
the color-suppressed deacy amplitude relative to the factorizable one, ranges
from $-{1\over 3}$ to $-0.60$ in the decays from $D\to
\bar{K}\pi$ to $D^+\to \phi\pi^+,~D\to\bar{K}^*\pi$.
By contrast, the variation of $a_2$ in $B\to\psi K,~B\to D(D^*)\pi(\rho)$
is negligible and nonfactorizable terms $\chi_2(B\to\psi K),~\chi_2(B\to
D^{(*)}\pi(\rho))$ are of order 10\% with a positive sign. The pattern
for the relative magnitudes of nonfactorizable effects
\be
|\chi(B\to PP,VP)|<|\chi(D\to PP)|<|\chi(D\to VP)|  \non
\en
is thus well established. This means that nonperturbative soft gluon effects
become more important when the final states are less energetic, allowing more
time for final-state interactions. This explains why $a_2$ is class ($PP$ or
$VP$ mode) dependent in charm decay, whereas it stays fairly
stable in $B$ decay.

Taking factorization as a benchmark,
we see that the nonfactorizable terms necessary for describing
nonleptonic $D$ and $B$ decays are in opposite directions from
the factorization framework. On the one hand, the leading $1/N_c$
expansion, which amounts to a universal $\chi=-{1\over 3}$, improves the naive
factorization method for charm decays. On the other hand, the naive
factorization hypothesis works better than the large-$N_c$ assumption
for $B$ decays because nonfactorizable effects are small, being of
order 10\%. The fact that $\chi$ is positive makes it even more clear that the
large-$N_c$ approach cannot be extrapolated from $D$ to $B$ physics.
Theoretically, the next important task for us is to understand why
$\chi$ is negative in $D$ decay, while it becomes positive in $B$ decay.

    As for $M\to VV$ decay, {\it a priori} effective $a_{1,2}$ cannot be
defined since, as pointed out by Kamal and Sandra, its amplitude (factorizable
and nonfactorizable) involves three independent Lorentz scalars,
corresponding to $S$, $P$ and $D$ waves. This turns out to be a nice
trade-off for solving the puzzle with the large longitudinal polarization
fraction $\Gamma_L/\Gamma$ observed in $B\to\psi K^*$, which cannot be
accounted for by the factorization hypothesis or by nonfactorizable effects
weighted in the same way in all three partial waves, namely $A_1^{nf}/
A_1=A_2^{nf}/A_2=V^{nf}/V$. A large $\Gamma_L/\Gamma$ can be achieved if
$B\to\psi K^*$ is almost all $S$-wave, implying that nonfactorizable
contributions are dominated by the $S$-wave.
The same assumption is also needed for
understanding the ratio ${\cal B}(B^-\to D^{*0}\rho^-)/{\cal B}(\bar{B}^0\to
D^{*+}\rho^-)$. We found that all nonfactorizable terms
$A_1^{nf}/A_1^{BK^*},~A_1^{nf}/A_1^{B\rho},~A_1^{nf}/A_1^{BD^*}$ are
of order 10\% consistent with $\chi_2(B\to D(D^*)
\pi(\rho))$ and $\chi_2(B\to\psi K)$.

   Surprisingly, the assumption of $S$-wave dominated nonfactorizable
effects is not operative in $D\to\bar{K}^*\rho$ decay, which exhibits again
another disparity between $B$ and $D$ physics. We found that
$A_2^{nf}/A_2$ should play a more pivotal role than $A_1^{nf}/A_1$ in
charm decay. We thus urge experimentalists to measure helicities
in the color- and Cabibbo-suppressed decay mode $D^+\to\phi\rho^+$ decay
to gain insight in the nonfactorizable effects in $D\to VV$ decay.
\vskip 2.0 cm
\centerline{\bf ACKNOWLEDGMENT}

    This work was supported in part by the National Science Council of ROC
under Contract No. NSC84-2112-M-001-014.

\pagebreak
\centerline{\bf REFERENCES}
\vskip 0.3 cm
\begin{enumerate}

\item See e.g. H.Y. Cheng, {\sl Int. J. Mod. Phys.} {\bf A4}, 495 (1989).

\item N. Deshpande, M. Gronau, and D. Sutherland, \pl {\bf 90B}, 431 (1980);
M. Gronau and D. Sutherland, \np {\bf B183}, 367 (1981).

\item H.Y. Cheng, \pl {\bf B335}, 428 (1994).

\item M. Wirbel, B. Stech, and M. Bauer, \zp {\bf C29}, 637 (1985).

\item A.J. Buras, J.-M. G\'erard, and R. R\"uckl, \np {\bf B268}, 16 (1986).

\item CLEO Collaboration, M.S. Alam {\it et al.,} \pr {\bf D50}, 43 (1994).

\item A.N. Kamal and A.B. Santra, Aberta Thy-31-94 (1994).

\item M. Gourdin, A.N. Kamal, and X.Y. Pham, \prl {\bf 73}, 3355 (1994).

\item R. Aleksan, A. Le Yaouanc, L. Oliver, O. P\`ene, and J.-C. Raynal,
DAPNIA/SPP/94-24, LPTHE-Orsay 94/15 (1994).

\item J.M. Soares, TRI-PP-94-78 (1994).

\item M. Bauer, B. Stech, and M. Wirbel, \zp {\bf C34}, 103 (1987).

\item Q.P. Xu, \pl {\bf B306}, 363 (1993).

\item C.A. Dominguez and N. Paver, \zp {\bf C41}, 217 (1988); A.A.
Ovchinnikov, {\sl Sov. J. Nucl. Phys.} {\bf 50}, 519 (1989); \pl {\bf B229},
127 (1989); V.L. Chernyak and I.R. Zhitnitski, \np {\bf B345}, 137 (1990);
S. Narison, \pl {\bf B283}, 384 (1992); V.M. Belyaev, A. Khodjamirian, and
R. R\"uckl, \zp {\bf C60}, 349 (1993); P. Colangelo, BARI-TH/93-152 (1993).

\item P. Ball, V.M. Braun, and H.G. Dosch, \pl {\bf B273}, 316 (1991); \pr
{\bf D44}, 3567 (1991); P. Ball, \pr {\bf D48}, 3190 (1993).

\item P. Colangelo and P. Santorelli, \pl {\bf B327}, 123 (1994).

\item K.C. Yang and W-Y.P. Hwang, NUTHU-94-17 (1994).

\item M. Witherell, in {\it Proceedings of the XVI International Symposium
on Lepton-Photon Interactions}, Ithaca, 10-15 August 1993, eds. P. Drell
and D. Rubin (AIP, New York, 1994).

\item L.L. Chau and H.Y. Cheng, \pl {\bf B333}, 514 (1994).

\item Particle Data Group, \pr {\bf D50}, 1173 (1994).

\item Mark III Collaboration, J. Adler {\it et al.,} \pl {\bf B196}, 107
(1987).

\item H.Y. Cheng and B. Tseng, IP-ASTP-21-94, hep-ph/9409408 (revised) (1994).

\item I. Halperin, TECHNION-PHYS-94-16 (1994).

\item B. Blok and M. Shifman, \np {\bf B389}, 534 (1993).

\item ARGUS Collabotation, H. Albrecht {\it et al.,} DESY 94-139 (1994).

\item CDF Collaboration, FERMILAB Conf-94/127-E (1994).

\item M. Gourdin, Y.Y. Keum, and X.Y. Pham, PAR/LPTHE/95-01 (1995).

\item C.E. Carlson and J. Milana, WM-94-110 (1994).

\item Mark III Collaboration, D. Coffman {\it et al.,} \pr {\bf D45}, 2196
(1992).

\item E691 Collaboration, J.C. Anjos {\it et al.,} \pr {\bf D46}, 1941 (1992).

\end{enumerate}
\end{document}